\documentclass[twocolumn,showpacs,preprintnumbers,amsmath,amssymb]{revtex4}
\usepackage{graphicx}
\usepackage{dcolumn}
\usepackage{bm}
\usepackage{sidecap}

\newcommand{\comment}[1]{}

\begin{document}


\title{The influence of photon angular momentum on ultrafast spin dynamics in Nickel}

\author{F. Dalla Longa}
\email[Electronic mail: ]{F.Dalla.Longa@tue.nl}
\author{J.T. Kohlhepp}
\author{W.J.M. de Jonge}
\author{B. Koopmans}

\affiliation{Department of Applied Physics and Center for
NanoMaterials, Eindhoven University of Technology, P.O.~Box~513,
5600 MB Eindhoven, The Netherlands}

\date{\today}

\begin{abstract}

The role of photon angular momentum in laser-induced demagnetization of Nickel thin films is investigated by means of pump-probe time-resolved magneto-optical Kerr effect in the polar geometry. The recorded data display a strong dependency on pump helicity during pump-probe temporal overlap, which is shown to be of non-magnetic origin. By accurately fitting the demagnetization curves we also show that demagnetization time $\tau_M$ and electron-phonon equilibration time $\tau_E$ are not affected by pump-helicity. Thereby our results do not support direct transfer of angular momentum between photons and spins to be relevant for the demagnetization process. This suggests, in agreement with the microscopic model that we recently presented, that the source of angular momentum could be phonons or impurities rather than laser photons as required in the microscopic model proposed by Zhang and H\"{u}bner.

\end{abstract}


\maketitle
Since the discovery by Beaurepaire et al. that excitation by femtosecond
laser pulses can induce a demagnetization in a Nickel thin film on a
sub-picosecond time scale \cite{baurepaire}, laser induced
magnetization dynamics received a growing attention \cite{Hohlfeld1997, Scholl1997, KoopPRL2000, Kirschner2000, Guidoni2002, Beaurepaire2004, bigot2005}. The
possibility of optically manipulating spins on such an ultrafast
time scale offers, indeed, many potential applications in
technology, e.g. in magneto-optical recording industry. Beside the
technological relevance, research in this field is motivated by
scientific interest, the microscopic mechanisms that lead to
ultrafast magnetization response being not yet fully understood.

Recently we presented a microscopic model that successfully explains the demagnetization process in terms of phonon- or impurity-mediated Elliot-Yafet type electron-electron spin-flip scattering, phonons and impurities providing the required transfer of angular momentum to the spins. Interestingly, the model also predicts a simple relation between two apparently unrelated parameters, the demagnetization time $\tau_M$ (i.e. the time scale that characterizes the magnetization loss triggered by laser excitation) and the Gilbert damping $\alpha$ \cite{koop2005PRL}. Using a different approach, Zhang and H\"{u}bner (ZH) attempted to explain the demagnetization process as the result of the combined action of spin orbit coupling (SOC) and the interaction between spins and laser photons \cite{zh}. The authors disregard the role of phonons, motivated by the expectation that conventional scattering mechanisms lead to spin-lattice relaxation times of some tens of picoseconds, too slow to account for the observed ultrafast demagnetization. Therefore the ZH model requires direct transfer of angular momentum between laser photons and electron spins to be responsible for the demagnetization. It is the aim of this paper to experimentally investigate whether such a transfer does indeed play a major role in the demagnetization process.

\begin{SCfigure*}
\includegraphics[width=1.445\columnwidth]{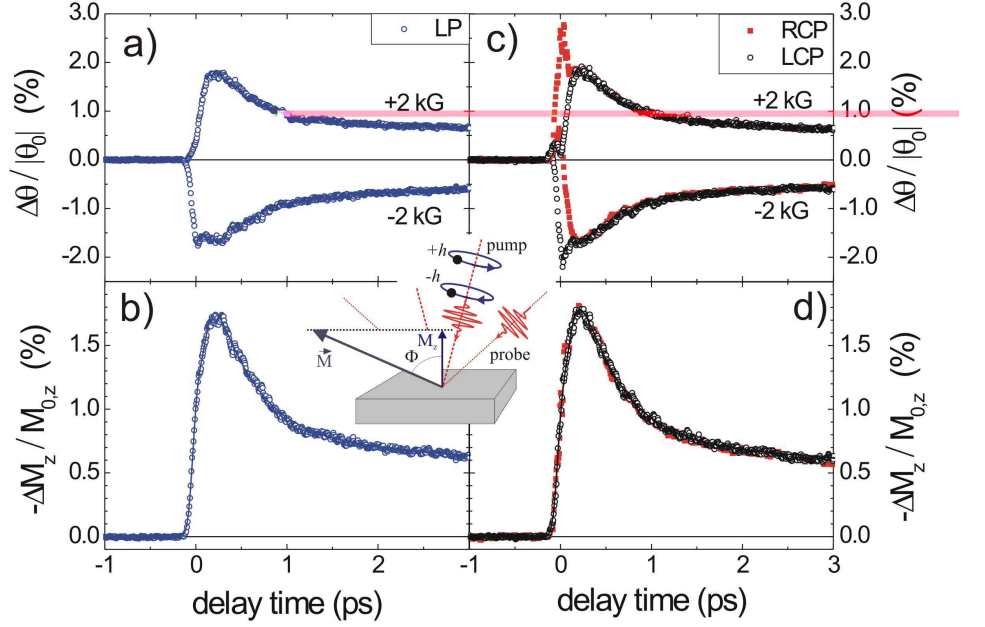}

\caption[fig1]{\label{fig:trmoke}(color online) Typical TR-MOKE response to (a) linearly polarized light (LP) pumping and (c) right (open circles) and left (full squares) circularly polarized light (CP) pumping, for an out of plane applied field of $\pm 2$kG. (b) genuine magnetization response to LP pumping obtained by averaging the curves in (a); (d) genuine magnetization response to right (open circles) and left (full squares) CP pumping obtained by averaging the corresponding right and left CP curves in (c). The solid lines in (b) and (d) are fits to the data using eq. \ref{eq:fit}. Inset: schematic representation of the experiment; the canted magnetization forms an angle $\Phi$ with the normal to the surface, CP pump photons carry a whole quantum of angular momentum $\pm \hbar$, probe pulses are sensitive to $M_z$.}

\end{SCfigure*}

In order to accomplish our goal, we performed optical pump-probe experiments on a Nickel thin film,
in which we make use of circularly polarized pump
pulses in the polar geometry. The most interesting feature that
distinguishes this approach from conventional pump-probe experiments
(where pump pulses are linearly polarized or unpolarized) is
that one can directly transfer not only energy, but also angular
momentum from the laser field to the electrons.

The first to report on CP pumping polar experiments on Ni films
were Wilks et al. \cite{wilks, wilks2}. In their investigations the authors
found that the response to CP pumping is characterized by the presence of an
additional peak superimposed to the conventional demagnetization
curve, centered around 0 ps delay time. The extra peak corresponds
to pump-probe crosscorrelation and does not affect the shape of
the demagnetization curve on a longer time scale. The presence of
the peak is due to the contribution of the so called Specular Inverse
Faraday Effect (SIFE) and Specular Optical Kerr Effect (SOKE), and from the variation of the peak height
with pump helicity one can derive the non-vanishing elements
of the cubic susceptibility tensor. Although the authors state
that, except for the presence of the extra peak, the
demagnetization signal is not affected by pump helicity, no
quantitative analysis is carried out in their work to support this
statement. In particular, the presence of the extra peak could
hide subtle pump-helicity dependent differences in the
demagnetization curve affecting the demagnetization time
$\tau_M$.

The measurements presented in this paper, besides confirming Wilks' results,
quantitatively show that angular momentum transfer due to CP pumping does not
contribute to the magnetization loss of more than $\pm 0.01 \%$, nor it influences the demagnetization time.

The sample under investigation consists of a 10 nm thick Ni film sputtered on a SiO substrate and capped with 2 nm of Copper to prevent from oxidation. The thickness of the ferromagnet has been especially chosen to match the light penetration depth ($\sim15$ nm for Ni at a wavelenght of 785 nm) in order to uniformly heat up the film throughout its thickness. Pump and probe pulses have a temporal FWHM of 70 fs and are focussed onto the same 8 $\mu$m diameter spot on the sample through a high aperture laser objective, with a final fluence of 2 and 0.1 mJ/cm$^{2}$ respectively. The laser pulses hit the sample at almost normal incidence: in this polar geometry the probe pulses are mostly sensitive to the out of plane component of the magnetization, $M_z$. A 2 kG field applied perpendicular to the film surface leads to a canted magnetization state inducing a finite $M_z$, as depicted in figure \ref{fig:trmoke}(inset).

In the time-resolved magneto-optical Kerr effect (TR-MOKE) setup, a quarter waveplate inserted in the pump beam's path enables the tuning of pump helicity between right circularly polarized (RCP) and left circularly polarized (LCP), and intermediate states. The linearly polarized probe pulses
pass through a Photo-Elastic Modulator (PEM) before being focussed
onto the sample; the PEM modulates the polarization of the pulses
from right circular to left circular with a frequency $f_{PEM}$. After reflection off
the sample the pulses are sent to a photo-detector through another
polarizer crossed with the first; it can be
shown \cite{koopmans2000JAP} that in these conditions the 2$f_{PEM}$
component of the detected signal is proportional to the
laser-induced changes of the Kerr rotation, $\Delta \theta$. In the time-resolved magnetization modulation spectroscopy (TIMMS) setup, one modulates the helicity of the pump pulses with a PEM while probe pulses are linearly polarized; therefore the differences between the responses to RCP and LCP pumping are directly detected. A more thorough description of TIMMS technique can be
found in the literature \cite{timms}; in particular it can be
shown that the 1$f_{PEM}$ component of the detected signal is
proportional to $\Delta \theta$. In
conclusion, with the TR-MOKE experiments we set the pump
polarization and measure the induced time-dependent
demagnetization, with the TIMMS experiments we modulate the pump
polarization and measure the time-dependent differences between
the demagnetization induced by RCP pumping and LCP pumping.

Before presenting the results, let us briefly explain the idea behind our experiment. Considering that the macroscopic magnetization is given by the space average of the atomic magnetic moments $\vec{\mu}=\mu_B(\vec{L}_e+g\vec{S}_e)$, where $\mu_B$ is the Bohr magneton, $\vec{L}_e$ and $\vec{S}_e$ are the orbital and spin part of electron momentum respectively, and $g\approx2$ for Nickel, it is clear that the
observed demagnetization is due to a \emph{laser induced momentum
transfer} from the electrons to ``somewhere else''. The allowed
demagnetization channels are given by conservation of total
angular momentum $\vec{J}=\vec{L}_e+\vec{S}_e+\vec{L}_{ph}+\vec{L}_{latt}$,
where $\vec{L}_{ph}$ and $\vec{L}_{latt}$ represent the angular
momentum carried by the photons and the phonons involved in the
process respectively. Therefore a demagnetization can happen due
to (i) exchange between orbital and spin part of electrons'
momentum through SOC, to (ii) momentum transfer from electrons to
the laser field and to (iii) momentum transfer from electrons to
the lattice. According to the ZH model interactions with the
lattice should be disregarded and only channels (i) and (ii) should be
considered as responsible for the demagnetization. Therefore in the framework of the ZH model the magnetization
response should strongly depend of the polarization of the pump
pulses. Since CP photons carry a whole
quantum of angular momentum $\pm\hbar$ along (RCP) or
opposite to (LCP) the direction of light propagation, transfer of angular momentum should induce a
demagnetization \emph{only} when photons helicity and magnetization are antiparallel, while the magnetization should
actually increase when they are parallel.

Our experimental configuration, in which $\vec{M}$ is only partially canted out of the sample plane, could seem inconvenient with respect to using a sample with perpendicular anisotropy. However our approach has the advantage that we can investigate the influence of pump helicity not only on demagnetization effects (i.e. affecting the modulus of $\vec{M}$), but also on orientational effects (i.e. affecting the canting angle $\Phi$). This is particularly interesting since recent experiments on garnet films showed that direct momentum transfer between photons and spins can non-thermally excite a coherent magnetization precession \cite{fredrik}. However no such effects were ever observed in our measurements, suggesting that in Nickel the response to laser excitation is dominated by thermal processes.

Let us now focus on our TR-MOKE experiments. The result of a
standard experiment, i.e. using linearly polarized light pumping, is presented in figure
\ref{fig:trmoke}(a), where the transient Kerr rotation normalized to its static value, $\Delta\theta/ |\theta_0|$, is plotted . When an out of plane field of 2 kG is applied,
the Kerr rotation displays a maximum at $\sim250$ fs
after laser excitation; if we reverse the field we see that the
same qualitative response with opposite sign is obtained. The
genuine magnetic response, $\Delta M_z / M_{0,z}$ (where $M_{0,z}$ represents the static value of $M_z$), is proportional to $\Delta\theta^+-\Delta\theta^-$,
the difference between the two Kerr rotation transients corresponding to opposite field directions, i.e. we are taking
into account only the part of the signal that changes sign upon
magnetization reversal; this is shown in figure \ref{fig:trmoke}(b).

The final dataset can be fitted with the following function:
\begin{eqnarray}
-\frac{\Delta M_z (t)}{M_{0,z}} & = & \Bigg[ \Bigg( A_1 F(\tau_0, t) - \frac{(A_2 \tau_E - A_1\tau_M)e^{-\frac{t}{\tau_M}}}{\tau_E - \tau_M} - {}
\nonumber\\
& & {} - \frac{\tau_E(A_1 - A_2)e^{-\frac{t}{\tau_E}}}{\tau_E-\tau_M}\Bigg)\Theta(t)+ A_3
\delta(t) \Bigg]\star {}
\nonumber\\
& & {} \star \Gamma(t) \label{eq:fit}
\end{eqnarray}
\noindent
where $\Gamma(t)$ is the Gaussian laser pulse, $\star$ represents
the convolution product, $\Theta(t)$ is the step function
and $\delta(t)$ is the Dirac delta function. The fitting function
of equation \ref{eq:fit} is derived from a solution to the so called three temperature model (3TM) in
the limit of low laser fluence, neglecting spin specific heat and
assuming an instantaneous rise of the electron temperature upon
laser excitation \cite{bert_rev}. The 3TM provides a phenomenological description of the demagnetization process in terms of energy redistribution among electrons, phonons and spins upon laser excitation (see also \cite{baurepaire}). The constant $A_1$ represents
the value of $-\Delta M_z/M_{0,z}$ after equilibrium between electrons,
spins and lattice is restored, if heat dissipation through the
substrate is neglected. The latter is described by the function $F(\tau_0, t)$, which depends on the specific features of the sample under investigation (in particular the conductive properties of the substrate) and is characterized by the time-scale $\tau_0 \gg \tau_E, \tau_M$. In our case the data are well described by a $1/\sqrt{t}$ behavior, i.e. $F(\tau_0, t)=\frac{1}{\sqrt{\frac{t}{\tau_0} + 1}}$. The constant $A_2$ is proportional to
the initial electron temperature rise, and determines the height of
the maximum of the curve. The constant $A_3$ represents the magnitude of optical artifacts taking place during pump-probe temporal overlap that can be well described by a delta function. The most important parameters are $\tau_E$
and $\tau_M$; the first describes the timescale of electron-phonon (e-p)
interaction (typically $\sim 450$ fs) that equilibrates the electron with the phonon system, the latter describes the time-scale of the
magnetization loss (typically $\sim 100$ fs).

In the specific case of the data of figure \ref{fig:trmoke}(b) we
obtained: $A_1=0.8 \%$, $A_2=2.33 \%$, $A_3=0.08 \%$, $\tau_M=73$ fs \cite{note1}, $\tau_E=440$ fs and $\tau_0=4.2$ ps. In the following sections we will only focus on the main timescales $\tau_E$ and $\tau_M$.

\begin{figure}[t,b]

\begin{center}
\includegraphics[width=1.0\columnwidth]{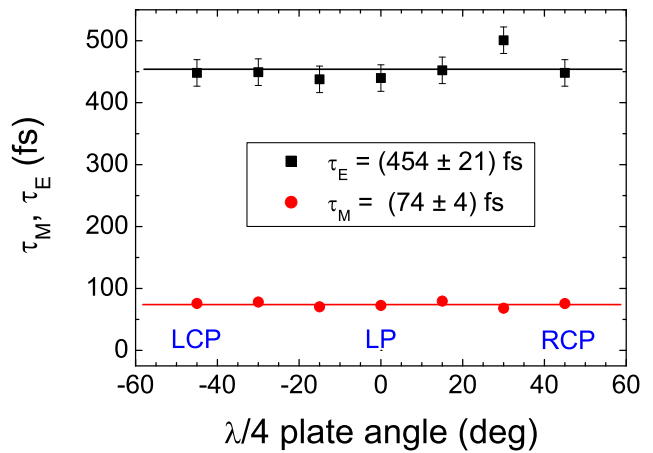}
\end{center}
\caption[fig2]{\label{fig:times} (color online) Demagnetization time $\tau_M$ (circles) and electron-phonon equilibration time $\tau_E$ (squares) against the orientation of the $\lambda/4$ plate: the values are nicely scattered around averages of 74 and 454 fs respectively (solid lines) showing no dependency on pump helicity. The error bars (not visible for the circles) are the standard deviations.}

\end{figure}


In figure \ref{fig:trmoke}(c) the demagnetization following RCP and
LCP light pumping is presented. As already reported in
\cite{wilks, wilks2}, when the system is pumped with CP light an
additional peak appears at 0 ps delay, superimposed to the usual
response. The extra peak does not change sign upon magnetization reversal, while
being strongly dependent on pump helicity, i.e. it changes sign
when pump helicity is inverted. The origin of the peak lies in the so called SIFE and SOKE contribution: in a simplified picture, CP photons transfer their angular momentum to the electronic orbits and the enhanced orbital momentum is then sensed by the probe beam.

Besides the presence of the additional peak, we notice that a
demagnetization is \emph{always} observed, independently of pump
helicity, even when photons angular momentum and
magnetization are parallel. This shows that direct transfer of
angular momentum between laser field and spins is not the main
mechanism giving rise to the demagnetization process, in contrast
with the predictions of the ZH model. At this point one can wonder
(i) if subtle differences between the response to RCP and LCP
light can be detected, and (ii) if
the time-scale of ultrafast demagnetization is affected by the
presence of the extra peak, i.e. if pump helicity influences
$\tau_M$.

In order to address point (ii) we proceed as in the linear case by
subtracting the two signals obtained at opposite fields: as it can be
seen in figure \ref{fig:trmoke}(d), the two curves overlap showing
no evident difference. In order to carry out a more quantitative
analysis we repeated the procedure for different values of
pump helicity, and fitted the resulting curves with equation
\ref{eq:fit}. The obtained values of $\tau_M$ and $\tau_E$ are
plotted as function of pump helicity in figure \ref{fig:times}: the
data are nicely scattered around average values $\bar{\tau}_M=(74\pm4)$ fs
and $\bar{\tau}_E=(454 \pm 21)$ fs, and show no
measurable dependency on pump helicity. This is a direct proof
that the time scales of demagnetization and e-p equilibration are
\emph{not} set by direct angular momentum transfer from the laser
field to the electron spins.

\begin{SCfigure}
\includegraphics[width=0.67\columnwidth, height=0.5\columnwidth]{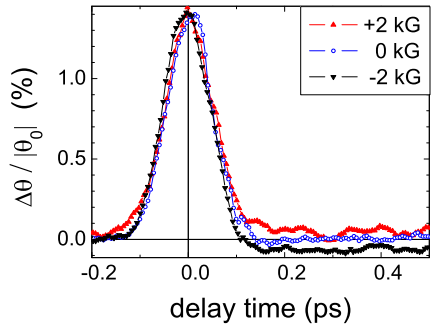}
\caption[fig3]{\label{fig:timms} (color online) TIMMS measurements: the field dependent signal after the SIFE/SOKE peak is due to a correlation between pump helicity and intensity (lines are guides to the eye).}
\end{SCfigure}

As for point (i), as we argued before in \cite{KoopPRL2000}, because of quenching of SOC, transfer of angular momentum to the spin degree of freedom is expected to be extremely small, of the order of $\frac{\Delta M_{phot}}{M_0} \sim \pm0.01\%$, comparable with the noise level in figure \ref{fig:trmoke}. Nevertheless, absorption of a circularly polarized photon leads to coherent transfer of angular momentum to the orbital component of the excited electronic state. Our data show, however, that after dephasing and re-establishing of the ground state ratio of spin and orbital momentum in the magnetic state, no significant transfer to the electronic system (magnetic moment) is left. In order to detect such a subtle difference a TIMMS modulation scheme can be
adopted. A typical dataset obtained from TIMMS measurements is
plotted in figure \ref{fig:timms}, the different curves
corresponding to different applied field values. The peak around 0
ps delay is once again the SIFE/SOKE contribution, and it is
independent of the applied field as one would expect from
the TR-MOKE experiments. After the SIFE/SOKE peak, the signal goes back to zero if no field is applied, while stabilizing to a small, though finite, value of $\pm 0.05\%$ when an out of plane field of $\pm 2$ kG is applied. If this subtle contribution came from a genuine difference in the response to RCP and LCP pumping due to a change in $\Delta M_z$ or $\Phi$ induced by direct transfer of angular momentum, one would expect it not to change sign upon field reversal. Therefore we conjecture that this small contribution is actually due to a finite correlation between pump helicity and pump intensity, due to the not exactly perpendicular incidence and to the presence of mirrors between the PEM and the sample. Therefore the TIMMS measurements support the picture of a photon contribution of less than $\pm 0.01\%$.

In conclusion we investigated the ultafast spin dynamic response
to CP laser light excitation in a Ni thin (10 nm) film by means of
TR-MOKE and TIMMS, aiming at a quantitative estimate of the contribution of photons angular momentum to laser-induced demagnetization. The analysis of the
data showed that the typical timescales involved in a
demagnetization experiment are not affected by the polarization of
the pump pulse; in particular we determined a demagnetization time
$\tau_M = (74\pm4)$ fs and an e-p equilibration time $\tau_E = (454\pm21)$ fs. Our results not only constitute an experimental proof that photon contribution to ultrafast demagnetization in Nickel is actually too weak to account for the observed effects, $\Delta M_z/M_{0,z}\approx 2\%$ in the case of figure \ref{fig:trmoke}, but also exclude that angular momentum transfer as required by the ZH model is the main mechanism leading to sub-picosecond demagnetization. Going back to angular momentum conservation in laser-induced demagnetization processes, one can conclude that a phonon-based microscopic mechanism should be considered. As we recently reported, both analytical \cite{koop2005PRL} and numerical \cite{koop+harm, MRSproc} calculations based on a
phonon-mediated electron-electron spin-flip scattering process
can predict the correct time scales.

The authors acknowledge W. H\"{u}bner for fruitful discussions.
The work is supported in part by the European Communities Human
Potential Programme under contract number HRPN-CT-2002-00318
ULTRASWITCH.


\begin{references}
\bibitem{note1}{We would like to remark that the same dataset would yield $\tau_M=135$ fs if empirically fitted by $[(A_1 F(\tau_0, t)+A_2 e^{-t/\tau_E})(1-e^{-t/\tau_M})\Theta(t)+A_3\delta(t)]\star \Gamma(t)$ as used in other works instead of equation \ref{eq:fit}.}
\end{references}
\end{document}